\documentclass[amsmath,longbibliography,superscriptaddress,reprint]{revtex4-2}

\usepackage{hyperref}
\usepackage{graphicx}
\usepackage{siunitx}

\begin{document}

\title{Deterministic Formation of Single Organic Color Centers in Single-Walled Carbon Nanotubes}

\author{Daichi Kozawa}
\email{kozawa.daichi@nims.go.jp}
\affiliation{Quantum Optoelectronics Research Team, RIKEN Center for Advanced Photonics, Wako, Saitama 351-0198, Japan}
\affiliation{Nanoscale Quantum Photonics Laboratory, RIKEN Pioneering Research Institute, Wako, Saitama 351-0198, Japan}
\affiliation{Research Center for Materials Nanoarchitectonics, National Institute for Materials Science, Tsukuba, Ibaraki 305-0044, Japan}

\author{Yuto Shiota}
\affiliation{Nanoscale Quantum Photonics Laboratory, RIKEN Pioneering Research Institute, Wako, Saitama 351-0198, Japan}
\affiliation{Department of Applied Physics and Physico-Informatics, Keio University, Yokohama, Kanagawa 223-8522, Japan}

\author{Yuichiro K. Kato}
\email{yuichiro.kato@riken.jp}
\affiliation{Quantum Optoelectronics Research Team, RIKEN Center for Advanced Photonics, Wako, Saitama 351-0198, Japan}
\affiliation{Nanoscale Quantum Photonics Laboratory, RIKEN Pioneering Research Institute, Wako, Saitama 351-0198, Japan}

\begin{abstract}
  Quantum light sources using single-walled carbon nanotubes show promise for quantum technologies but face challenges in achieving precise control over color center formation. Here we present a novel technique for deterministic creation of single organic color centers in carbon nanotubes using \textit{in-situ} photochemical reaction. By monitoring discrete intensity changes in photoluminescence spectra, we achieve precise control over the formation of individual color centers. Furthermore, our method allows for position-controlled formation of color centers as validated through photoluminescence imaging. We also demonstrate photon antibunching from a color center, confirming the quantum nature of the defects formed. This technique represents a significant step forward in the precise engineering of atomically defined quantum emitters in carbon nanotubes, facilitating their integration into advanced quantum photonic devices and systems.
\end{abstract}

\maketitle

\section{Introduction}
Quantum light sources capable of emitting single photons on demand are essential for emerging quantum technologies~\cite{Aharonovich2016}, enabling advances in communication~\cite{Gisin2007}, computing~\cite{DiVentra2013}, and sensing applications~\cite{Degen2017}. Notable developments toward these applications include single-photon emission demonstrated in various solid-state systems such as nitrogen-vacancy centers in diamond~\cite{Kurtsiefer2000}, semiconductor quantum dots~\cite{Michler2000}, and defects in two-dimensional materials~\cite{Koperski2015,Kozawa2023, Tran2016}. Particularly desirable are quantum light sources that operate at room temperature and within the telecom wavelength range~\cite{He2017,Ishii2017}, while material limitations and emission efficiency constraints need to be considered.

Single-walled carbon nanotubes (SWNTs) have emerged as a promising platform for quantum light sources~\cite{Ishii2017}, owing to their unique one-dimensional structure~\cite{Saito2000}, exceptional optical properties~\cite{Avouris2008, Kozawa2024}, and compatibility with telecom wavelengths~\cite{Weisman2003, Zaumseil2022, He2017, Ishii2015}. The introduction of organic color centers in SWNTs has further enabled single-photon emission at room temperature~\cite{He2017} with large tunability of the emission energy~\cite{Kwon2016}. Nevertheless, one of the critical challenges is the deterministic control over the formation of these quantum defects, particularly the number of defects introduced and their spatial positioning within single carbon nanotubes~\cite{Dou2021,Huang2020}. Existing methods often lack the capability or the precision required, limiting their potential for scalable quantum photonic applications~\cite{Ishii2018}.

In this work, we develop a technique for the deterministic creation of single organic color centers using \textit{in-situ} photochemical reaction. As nanotubes are functionalized, discrete photoluminescence (PL) intensity changes corresponding to the formation of individual color centers are observed. By stopping the reaction upon detecting the discrete increase, we are able to deterministically create single color centers. Statistical analysis of PL spectra from individual color centers is conducted to obtain quantitative insight into the distribution of color center types. In addition, this technique allows for position-controlled formation of color centers which is validated by PL imaging. Furthermore, we observe photon antibunching from a color center by performing a photon correlation measurement, showing that single quantum defects can be formed using this technique. 

\section{Results and discussion}

\begin{figure*}[t]
  \includegraphics{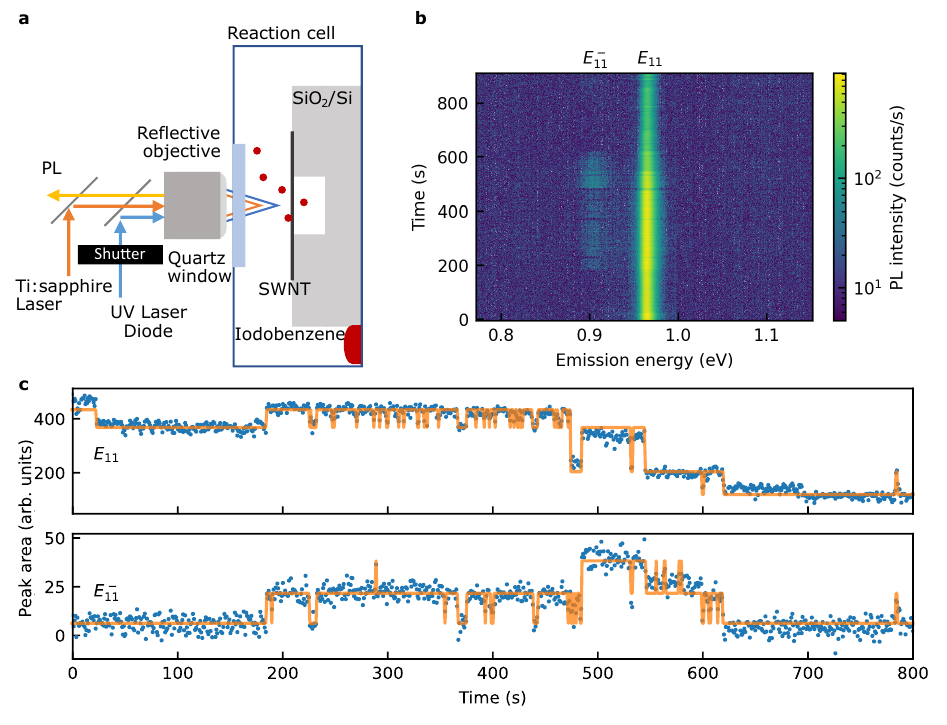}
  \caption{
      \label{fig1}
      (a) A schematic of an experimental setup for the \textit{in-situ} functionalization. (b) A time-trace map of PL spectra for a (9,7) SWNT excited with 1.59~eV and 20~\textmu W. (c) Temporal profiles of emission peaks $E_{11}$ and $E_{11}^{-}$ extracted from the time trace (b) by spectrally integrating the intensity with a bin width of 20~meV at each time point. The solid lines are fits by the Gaussian mixture model. The shutter for the UV laser is open after $t = 0$~s in panels (b, c). 
  }
\end{figure*}

Air-suspended SWNTs are grown across trenches on Si substrates by chemical vapor deposition (CVD)~\cite{Ishii2015, Ishii2017, Ishii2019}, with the nanotube density carefully controlled via growth parameters to allow for single tube measurements. The substrate is then placed inside a sealed reaction cell, where a droplet of iodobenzene is deposited next to the substrate and left for 10~min to saturate the chamber with vapor. The cell is subsequently mounted on a motorized three-dimensional feedback stage, which enables precise spatial targeting of individual SWNTs. Local functionalization is performed via a vapor-phase photochemical reaction~\cite{Kozawa2022} by focusing an ultraviolet (UV) laser through a quartz window onto a selected SWNT (Fig.~\ref{fig1}a). Initially, the UV laser is blocked by a shutter, and PL spectra are acquired over a time period of 5~s using Ti:sapphire laser excitation. At $t = 0$~s, the shutter is opened and UV irradiation begins while PL spectra continue to be recorded.

During the photochemical reaction of a (9,7) SWNT, PL spectra show temporal changes (Fig.~\ref{fig1}b). The pristine nanotube initially exhibits only $E_{11}$ emission, while an additional peak labeled $E_{11}^{-}$ emerges at $t = 190$~s, indicating that functionalization has occurred. Spectral diffusion or broadening of the emission peaks is insignificant in most nanotubes during the reaction, enabling us to focus on the emission intensity for further analysis. We extract temporal profiles of the emission intensity by spectrally integrating the PL spectra of each relevant peak in the time trace map (Fig.~\ref{fig1}c; see Methods for details). Notably, we observe discrete intensity steps in both $E_{11}$ and $E_{11}^{-}$ emission, which can be attributed to the formation of individual organic color centers. For the $E_{11}$ emission, most of these discrete steps correspond to intensity reductions, likely caused by the introduction of color centers or quenching sites. While many of the intensity steps occur synchronously in both $E_{11}$ and $E_{11}^{-}$, some steps in the $E_{11}$ intensity occur without changes in the $E_{11}^{-}$ intensity, suggesting the formation of quenching sites independent of organic color center formation.

The time traces can be well reproduced by the Gaussian mixture model (GMM)~\cite{Bandyopadhyay2021}, which provides a statistical framework for distinguishing discrete emissive states. We combine GMM with the Akaike information criterion to minimize overfitting while ensuring an accurate representation of the data, allowing for an objective determination of the optimal number of formed color centers. The GMM validates that these steps are not random intensity fluctuations but discrete, quantized transitions, suggesting the individual creation of color centers.

\begin{figure*}[t]
  \includegraphics{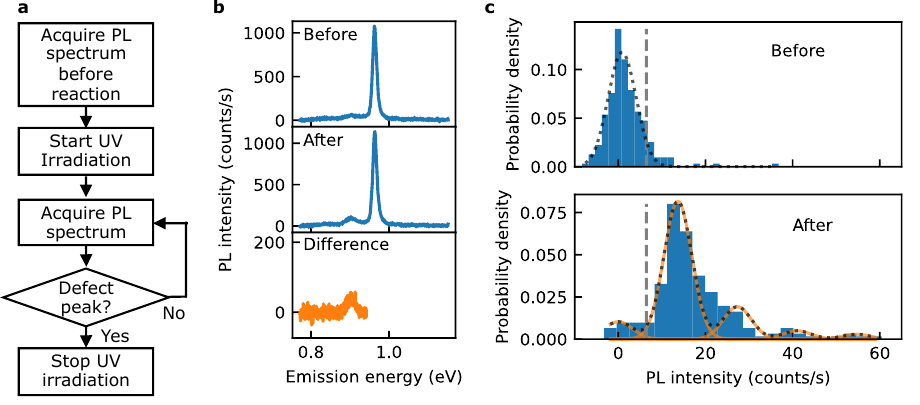}
  \caption{(a) A flow chart of the algorithm for a single color center formation. (b) Spectral difference before and after reaction of a (9,7) SWNT. (c) Peak probability densities of the $E_{11}^{-}$ intensity before and after the reaction. The black dotted lines are sum of the Gaussian fit with Eq.~\ref{eq1}, the orange lines are individual Gaussian components, and the vertical gray broken lines indicates 1$\sigma$ of the intensity distribution before the reaction. The functionalization is conducted with a UV laser power of 5~nW, while PL spectra are taken with an excitation energy of 1.59~eV and a power of 100~\textmu W. The PL intensity is obtained by computing the mean intensity within a 12~meV window centered at each emission peak.}
  \label{fig2}
\end{figure*}

Using the discrete intensity steps observed in the PL spectra, we can control the photochemical reaction~\cite{Wu2018} to reliably form single color centers. Figure~\ref{fig2}a summarizes the algorithm employed for this process, in which we first acquire a PL spectrum before initiating the reaction. After we start UV irradiation, the reaction is closely monitored in real time by repeatedly acquiring PL spectra every 0.5~s. In order to detect the emergence of color center emission, we compute the difference of the real time spectrum and the one taken before UV irradiation. The mean of the difference intensity is computed within an energy window of 12~meV below $E_{11}$, and a predefined threshold given by the root-mean-square intensity $\sigma$ of all the spectra before the reaction is used as a criterion to stop UV irradiation. This procedure is adopted for all subsequent preparation of color centers. In Fig.~\ref{fig2}b, a typical spectral difference before and after the reaction highlights the appearance of an emission peak, which clearly indicates color center formation. The calculation of the difference can cancel out the contribution of a phonon sideband around $E_{11}^-$ which is assigned to the out-of-plane transverse optical/out-of-plane transverse acoustic phonon modes at the $K$ point~\cite{Kozawa2024}. 

To characterize the effectiveness in creating color centers, we analyze the statistical significance of the intensity change in 274 PL spectra of individual functionalized (9,7) SWNTs. As shown in Fig.~\ref{fig2}c, the peak probability density of the $E_{11}^{-}$ emission intensity demonstrates a significant shift beyond 1.5$\sigma$ before the reaction. The histogram after the reaction shows a long tail towards high intensities, which may indicate the presence of multiple color centers. We therefore fit the PL intensity distribution using a multiple Gaussian function to gain insight into the formation of color centers. The distribution of the PL intensity $I$ is well described by a probability density function
\begin{equation}\label{eq1}
  P(I) = \sum_{j=0}^{n} a_j \exp\left(-\frac{(I - j \mu)^2}{2 \sigma^2} \right),
\end{equation}
where $n$ is the maximum number of color centers considered, $a_j$ represents the peak probability density for $j$ color centers created, and $\mu$ is the mean intensity of a single color center. The pre-reaction distribution is well described by $n=0$, while the post-reaction distribution requires $n=4$ for optimal fitting. When fitting the post-reaction PL intensity distribution, we use the $\sigma$ value obtained from the pre-reaction fit, assuming detector noise is the dominant source of the broadening. The fitted distribution reveals evenly spaced intensity clusters, which can be explained by the formation of $j=0$ through $4$ color centers.

The distribution obtained from the experiments shows no evidence of higher probabilities for multiple defect formation~\cite{Nutz2019,Kariuki1999,Wang2017}. This suggests that color center formation is not influenced by the presence of other color centers. If the initial radical attachment to the nanotube wall alters the local electronic and chemical environment, the reactivity of adjacent carbon atoms may be enhanced to favor successive formation. Direct structural characterization using scanning transmission electron microscopy or scanning tunneling microscopy would provide deeper understanding of the color center formation process.

The fraction of spectra with $j > 1$ is relatively small, being less than 23\% as observed in the intensity histogram (Fig.~\ref{fig2}c). The formation of multiple color centers can be further minimized by lowering the reaction rate, which will provide sufficient time to stop the reaction before multiple color centers are formed. Additionally, the probability for $j = 0$ can be suppressed by increasing the integration time of the PL spectra to reduce $\sigma$. Our \textit{in-situ} reaction and monitoring technique therefore allows for deterministic formation of single color centers on demand.

\begin{figure}[t]
  \centering
  \includegraphics{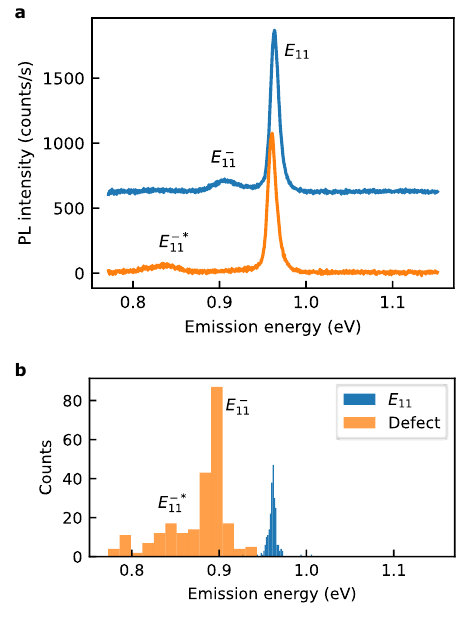}
  \caption{
    Statistical distribution of the emission energies after the functionalization of (9,7) SWNTs. (a) PL spectra displaying $E_{11}^{-}$ and $E_{11}^{-*}$ defect emission, where the spectrum with $E_{11}^{-}$ is vertically displaced. (b) Histograms of the peak center obtained with a fit by the Lorentz function. The functionalization is conducted with a UV laser power of 5~nW, while PL spectra are taken with an excitation energy of 1.59~eV and a power of 100~\textmu W. The bin width of the histograms are determined by Freedman Diaconis estimator. For analysis of reacted nanotubes, only spectra with an $E_{11}$ peak area of at least 60~eV$\cdot$counts/s and a linewidth between 10 and 80~meV are used.}
  \label{fig3}
\end{figure}

We now proceed to analyze the spectra acquired after the formation of the color centers and examine the statistical distribution of the emission energies to obtain insight into the defect types. Typical PL spectra shown in Fig.~\ref{fig3}a highlight the emission features from two distinct defect types $E_{11}^{-}$ and $E_{11}^{-*}$. The difference in the emission energy has been interpreted to arise from the binding configurations of functional groups~\cite{Gifford2019,Saha2018,He2017a}. We fit all collected PL spectra by a double Lorentzian function where the emission peak with an energy lower than $E_{11}$ is assigned to color center emission. Histograms of the emission energies show a distribution corresponding to the different defect types (Fig.~\ref{fig3}b). The main peak in the distribution underscores the reproducibility of defect formation and indicates preferential pathways in the photochemical reaction to form $E_{11}^{-}$ color centers. 

\begin{figure*}[t]
  \includegraphics{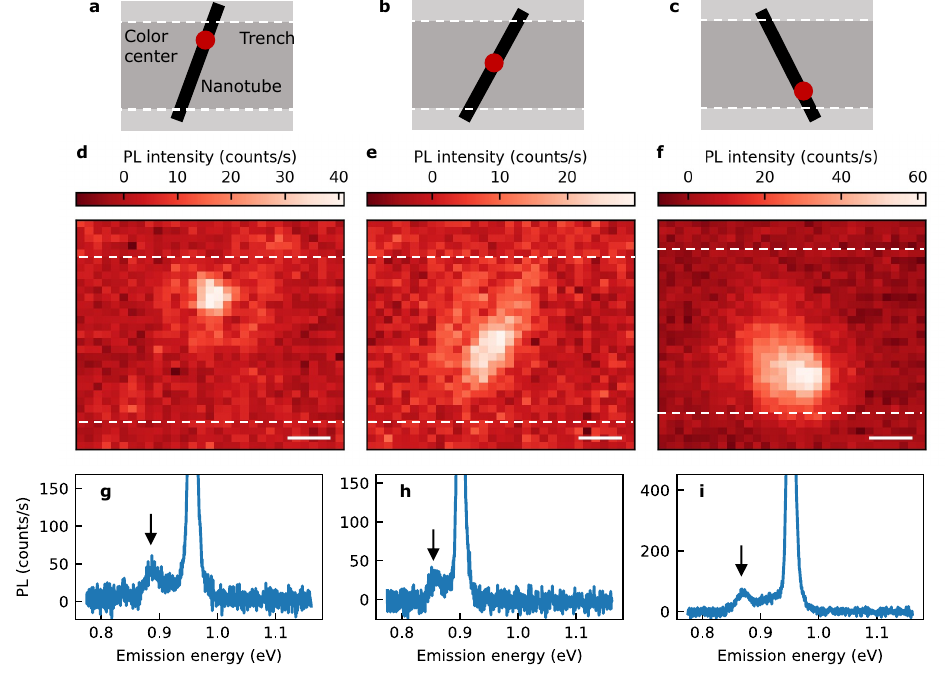}
  \caption{
      \label{fig4}
      Spatial control over color center formation in SWNTs. (a--c) Schematics of air-suspended SWNTs where target positions are indicated by red dots and trench edges are indicated by broken lines. (d--f) Excitation PL images of nanotubes functionalized at different positions where the images (d--f) are obtained with the spectrally integrated intensity within 4-meV windows at the color center emission peaks indicated by arrows in PL spectra (g--i) respectively. The functionalization is conducted with a UV laser power of 5~nW, while PL spectra are taken with an excitation energy of 1.59~eV and a power of 100~\textmu W. 
      }
\end{figure*}

In addition to controlling the number of color centers, we demonstrate position-controlled formation in individual SWNTs. By targeting specific regions of the nanotubes, we perform localized functionalization as depicted schematically in Fig.~\ref{fig4}a--c where the UV laser is focused at the top, middle, and bottom regions of the tubes. Locations of color centers are characterized by excitation imaging measurements where we scan over the nanotubes to excite and collect PL spectra. Intensity maps for the color center peaks (Fig.~\ref{fig4}d--f) are then constructed by spectrally integrating the intensity within energy windows at the color center emission peaks (Fig.~\ref{fig4}g--i). The images provide visual evidence of the position-controlled color center formation. We note that the size of the bright regions in the PL images differs from defect to defect, which does not necessarily imply differences in defect numbers or distributions since the intensity maps are not emission images but are obtained by scanning the excitation laser. The bright region is most localized in Fig.~\ref{fig4}d, which is attributed to formation of quenching defects in proximity to the bright color center. In comparison, a noticeable blur around the defect sites, visible in Fig.~\ref{fig4}e and Fig.~\ref{fig4}f, suggests the influence of exciton diffusion~\cite{Ishii2015,Moritsubo2010,Ishii2019}. The difference in sharpness between the images supports the interpretation that quenching defects and color centers are generated independently during the reaction.

\begin{figure}[t]
  \centering
  \includegraphics[width=7.6cm]{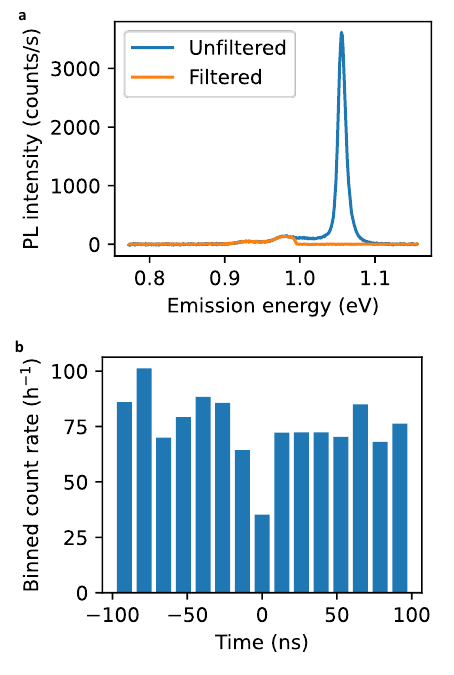}
  \caption{(a) PL spectra of a functionalized (11,3) SWNT showing an $E_{11}^-$ emission peak. The orange and blue spectra are taken with and without a long-pass filter having a cut-off energy of 0.992~eV, respectively. (b) Photon correlation showing $g^{(2)}(0) = 0.45$ taken with 200~nW, 1.59~eV excitation, and a 2-hour accumulation time.}
  \label{fig5}
\end{figure}

The capabilities for forming single color centers at desired locations provide an important step towards applications in quantum light sources~\cite{Li2021a}. To this end, we perform a photon correlation measurement using a Hanbury-Brown-Twiss setup under pulsed excitation which is a definitive test for single-photon emission, as it allows us to evaluate the photon antibunching behavior. Figure~\ref{fig5}a shows PL spectra of a (11,3) SWNT with and without a long-pass filter, where the filter isolates the $E_{11}^-$ color center emission from the $E_{11}$ emission. The photon correlation data presented in Fig.~\ref{fig5}b reveal a clear antibunching behavior characterized by a second-order correlation function value of $g^{(2)}(0) = 0.45$, indicating a significant suppression of multi-photon events compared to random photon emission and confirming the single-photon nature of the emission. 

In summary, we have developed \textit{in-situ} photochemical reaction technique that enables the deterministic formation of single organic color centers in air-suspended SWNTs. By monitoring PL spectra in real time, we have observed discrete intensity changes that correspond to the formation of individual color centers. The introduction of the color centers can then be precisely controlled by blocking UV irradiation upon the first detection of color center emission. Statistical analysis of PL spectra has revealed a preference for the formation of $E_{11}^{-}$ emitters, and spatial control over defect placement has been demonstrated via PL imaging. Furthermore, the photon correlation measurement has confirmed single-photon emission, establishing the quantum nature of the defect. Our deterministic functionalization technique offers the potential for on-demand use of quantum defects with desired emission energies, while providing capability for broader spectral coverage by the choice of chirality~\cite{Kozawa2022} and molecular precursors~\cite{Kim2018,Kwon2016}. This level of control paves the way for the development of atomically defined technology for scalable quantum photonic circuits, operating at room temperature within the telecom band.

\section*{Methods}
\subsection{Fabrication of Air-Suspended Carbon Nanotubes}
Our process for fabricating air-suspended single-walled carbon nanotubes (SWNTs) utilizes a combination of electron-beam lithography and dry etching techniques~\cite{Ishii2015}. We begin by employing these methods to create trenches on silicon substrates. The trenches are approximately 1.0-\textmu m deep and can be up to 4.0-\textmu m wide. Following this step, a secondary electron-beam lithography process is performed to define the catalyst areas in the vicinity of these trenches. We then spin-coat these areas with a Fe-silica catalyst that has been dispersed in ethanol. Excess catalyst is removed through a lift-off process, ensuring that it remains only within the predefined areas. Synthesis of the SWNTs occurs over these trenches with alcohol CVD~\cite{Ishii2019,Ishii2017}. The synthesis process is conducted under a flow of ethanol with a carrier gas mixture of argon and hydrogen at 800$^\circ$C for 1~minute. The result is air-suspended SWNTs positioned over the trenches, ready for further experiments.

\subsection{Micro-PL Measurements}
We conduct PL characterization using a home-built scanning confocal microscope~\cite{Ishii2015,Ishii2019}. For these experiments, we use a continuous-wave Ti:sapphire laser for excitation and a liquid-N$_2$-cooled InGaAs photodiode array attached to a 30-cm spectrometer for detection. The laser polarization is maintained perpendicular to the trenches, and the beam is focused using a reflective objective lens with a numerical aperture of 0.5 and a working distance of 7.8~mm. The 1/$e^2$ diameter of the focused beam is 1.32~\textmu m for an excitation energy of 1.59~eV. These diameters are characterized by performing PL line scans perpendicular to a suspended tube. The confocal pinhole defines the collection spot size, which is approximately 5.5~\textmu m in diameter. For PL imaging, we scan over a SWNT to collect PL spectra and construct intensity maps for color center emission by spectrally integrating the intensities of each peak. All PL spectra are taken at the center of the nanotubes except for the hyperspectral PL images. All measurements are carried out at room temperature.

\subsection{Formation of Organic Color Centers}
To functionalize the air-suspended nanotubes with iodobenzene as a precursor, we use vapor-phase reaction~\cite{Kozawa2022}. As-grown SWNTs on Si substrates are placed in a reaction cell with an inner volume of 7.4~mL. We introduce 20~\textmu L of iodobenzene to the bottom of the cell using a micropipette, and the cell is sealed in air. After 10~minutes to allow the cell to fill with iodobenzene vapor, we perform the reaction by irradiating the SWNTs with a 303-nm UV laser through the quartz window of the cell, which has a thickness of 0.5~mm. We control the UV laser exposure using a motorized shutter with a closing time of 8.0~ms, allowing precise on-off switching during the photochemical reaction. The UV laser beam is colinearly aligned with the Ti:sapphire laser beam, where the two beams are overlapped at a long-pass dichroic mirror just before the reflective objective lens. The polarization of the UV laser is kept parallel to the trenches. Following the reaction, we remove the samples from the cell and store them in dark for subsequent spectroscopic characterization.

\subsection{Analysis of PL Time Traces}
In the analysis of PL spectral time traces, the first step is to identify emission peaks by applying a peak-finding algorithm to a variance spectrum derived from a time trace of PL spectra. Once identified, the intensity of each emission peak is spectrally integrated using a bin width of 20~meV, and the resulting intensity is plotted as a function of time to construct temporal profiles. The PL intensity profiles are then fitted using the Gaussian mixture model, with the number of intensity levels determined based on the Akaike information criterion.

\subsection{Photon Correlation Measurement}
The photon correlation measurement is conducted using a Hanbury-Brown-Twiss setup with a 50:50 fiber coupler under laser excitation of 100-fs pulses from a Ti:sapphire laser operating at a repetition rate of 76~MHz~\cite{Fang2024,Li2021}. The excitation laser beam is focused onto the sample through a transmissive objective lens with a numerical aperture of 0.85. PL from the center of the nanotube is coupled via single-mode fibers to superconducting single-photon detectors, and data collection is performed using a time-correlated single-photon counting module.

\section*{Acknowledgement}
This work was supported in part by JSPS KAKENHI Grants No. JP24H01210, No. JP23K23161, and No. JP23H00262; JST ASPIRE Grant No. JPMJAP2310; the Canon Foundation; the Mitsubishi Foundation; and MEXT ARIM Grant No. JPMXP1222UT1136. We thank the Advanced Manufacturing Support Team at RIKEN for their technical assistance.

\end{document}